\documentclass[a4paper,11pt]{article}
\pdfoutput=1 % if your are submitting a pdflatex (i.e. if you have
\usepackage[utf8]{inputenc}
\usepackage{jcappub}

\usepackage{morefloats}

\newcommand{\uh}{UHECR}
\newcommand{\uhs}{UHECRs}
\newcommand{\aug}{AUG}
\newcommand{\ic}{IceCube}
\newcommand{\sbg}{SBGs}
\newcommand{\ns}{neutrinos}
\newcommand{\n}{neutrino}

\title{Are starburst galaxies a common source of high energy neutrinos and cosmic rays?}

\author[a]{Cecilia Lunardini,}
\author[b]{Gregory S. Vance,}
\author[c]{Kimberly L. Emig,}
\author[b]{Rogier A. Windhorst}

\affiliation[a]{Department of Physics, Arizona State University, Tempe, AZ 85287-1504 USA}
\affiliation[b]{School of Earth and Space Exploration, Arizona State University, Tempe, AZ 85287-1404 USA}
\affiliation[c]{Leiden Observatory, Leiden University,PO Box 9513, NL-2300 RA Leiden, the Netherlands}

\emailAdd{Cecilia.Lunardini@asu.edu}
\emailAdd{Gregory.S.Vance@asu.edu}
\emailAdd{emig@strw.leidenuniv.nl}
\emailAdd{Rogier.Windhorst@asu.edu}

\abstract{
 A recent analysis of cosmic ray air showers observed at the Pierre Auger Observatory indicates
 that nearby starburst galaxies (\sbg) might be the cause of $\sim10\%$ of the  Ultra-High-Energy Cosmic Ray flux at energies $E > 39$ EeV. Since high energy neutrinos are a direct product of cosmic ray interactions, we investigate \sbg\ as a possible source of some of the 
 ${\sim 10^{-2} - 1}$ PeV 
  \ns\ observed at IceCube. A statistical analysis is performed to establish the degree of positional correlation between the observed \ns\ and a set of 45 nearby radio- and infrared-bright \sbg.   Our results are consistent with no causal correlation. However, a scenario where $\sim 10\%$ of the 
High Energy Starting Events (HESE) in the detector
  are coming from the candidate \sbg\ is not excluded. The same conclusion is reached for 
  different data subsets, 
  as well as two different subsets of \sbg\ motivated by the Pierre Auger Observatory analysis. 
}


\begin{document}
\maketitle
\flushbottom

\section{Introduction}

The Ultra-High-Energy Cosmic Rays (\uhs) that the Earth constantly receives from space are the most energetic particles ever observed (see, e.g. \cite{Kampert:2012vi,LetessierSelvon:2011dy} for an introduction).  They are mainly hadrons (protons and/or atomic nuclei) with energies exceeding the EeV scale \cite{Linsley:1963km}.  A long-standing goal is to identify the 
astrophysical sources of
the \uhs\ and, ultimately, understand the acceleration mechanisms that take place there.

In the last five years, cosmic ray physics has entered the multi-messenger era, where cosmic ray and gamma ray data are being complemented by detections of gravitational waves \cite{Abbott:2016blz} and \ns. About $ 100$ \ns, with observed energies $ 0.02 -2$ PeV, have been detected by the kilometer-scale experiment \ic\ since 2013 \cite{Aartsen:2013bka}.  Most of these \ns\ have likely originated from cosmic rays, having been produced in the collision of cosmic rays with ambient protons or photons,  either in the sources themselves, or in the medium between the sources and the Earth. Considering that \ns\ propagate unabsorbed and undeflected over cosmological distances, they are ideal probes of the sites and origin of high energy particle acceleration. 

Ultimately, the definitive answer to the question of the origin of the \uhs\ and the high energy \ns\ will be given by an evidence of positional correlation of the observed particles with candidate sources. Therefore, searches for correlations are crucial, and intense multi-messenger searches are ongoing on this front. 
Recently, an analysis of the data of the Pierre Auger Observatory  \cite{Aab:2018chp}  
(we will refer to this paper as \aug\ from here on) 
showed an indication of positional correlation of the highest energy cosmic rays with Starburst Galaxies (\sbg), which are characterized by exceptionally high rates of star-formation.  Specifically, for the particles with (observed) energies $E > 39$ EeV, a model with 9.7\% of the \uh\ flux from nearby \sbg\ (and the remaining $90.3\%$ isotropic) was found to be favored, with $4\sigma$ significance, over a completely isotropic scenario.  
About $90\%$ of the  anisotropic flux was found to be attributable to four nearby \sbg: NGC 4945, NGC 253, M83, and NGC 1068.  
The \aug\ claim was checked in a new analysis of the Telescope Array (TA) data  \cite{Abbasi:2018tqo}. The results were inconclusive, being consistent with both the \aug\ anisotropy (within $1.4~\sigma$) and with the hypothesis of complete isotropy (within $1.1~\sigma$). 
An independent analysis of the Pierre Auger Observatory data, employing a joint fit of cosmic ray arrival directions and energy spectra,  reached conclusions that are broadly consistent with \aug\ \cite{Capel:2018cnf}. 
 The recent results by Auger and TA add to a broader picture, where preliminary indications of correlations between neutrinos and \uhs\ have emerged, in the context of cosmic ray hotspots possibly due to \sbg\ (see, e.g., \cite{Fang:2014uja,Aartsen:2015dml}, and \cite{Anchordoqui:2018qom} for a recent review). 

The Auger/TA results renew the interest in \sbg\ as possible cosmic ray and \n\ sources.   
Theoretically, they are well motivated (see e.g., \cite{Romero:2003tj,Loeb:2006tw,Stecker:2006vz,Murase:2013rfa,Tamborra:2014xia}). Indeed, they host episodes of extremely efficient star-formation, which causes high rates of core collapse supernovae. The resulting supernova ejecta propagate into the interstellar gas, producing shocks where cosmic ray acceleration and \n\ production takes place.
Observationally, positional coincidences between the \n\ arrival directions and nearby \sbg\ were noticed early on  \cite{Anchordoqui:2014yva}. In a statistical analysis, three of us \cite{Emig:2015dma} found an excess -- although not significant -- of coincidences compared to the prediction in absence of a causal relationship. Overall, statistical analyses of coincidences  \cite{Emig:2015dma,Moharana:2016mkl,Hooper:2018wyk} constrain the contribution of \sbg\ to tens of percent of the total  astrophysical \n\ flux, which is in agreement with arguments of consistency between the \ic\ data and gamma ray observations \cite{Murase:2015xka,Bechtol:2015uqb,Sudoh:2018ana} (see also \cite{Tamborra:2014xia,Chang:2014hua,Chang:2014sua}).  However, the situation of \n\ analyses is still open to including new data, and to including various uncertainties on the consistency argument (see, e.g., \cite{Xiao:2016rvd,Chakraborty:2016mvc,Linden:2016fdd,Palladino:2018bqf}).
Hence, an update in the light of the \aug\ result is in order.

 In this paper, such an update is presented.  There are two main elements of novelty. First, we obtain a new compilation of \sbg\ that extends the one used in the \aug\ paper to the southern hemisphere. This is especially important, considering that a large fraction of the \n\ data is located in that part of the sky.  Secondly,  we use the latest published \ic\ data to test for positional associations of the \ns\ with this expanded set of \sbg.  The results can be directly compared to those in \aug, and therefore they serve as a natural complement to it, adding one more piece of information to the general multi-messenger landscape. 
In sec. \ref{sec:data}, a description of the method and of the data used is given; our results are presented in sec. \ref{sec:res}, followed by a brief discussion in sec. \ref{sec:disc}.  More details on our \sbg\ compilation are available in Appendix \ref{app:catalogs}.

%%%%%%%%%%%%%%%%%%%%%%%%%%%%%%%%%%%%%
\section{Data and methodology}
\label{sec:data}

%%%%%%%%%%%%%%%%%%%%%%%%%%%%%%%%%%%%%
\subsection{Neutrino observations and SBG catalogs}
\label{subsec:data}

We consider the two most extended sets of \ic\ data that are publicly available.  The first is the  6-year data set of high-energy starting events (HESE) \cite{Aartsen:2014gkd,Kopper:2015vzf,Kopper:2017zzm}, which refers to candidate \n\ detections (``events") for which the \n\ interaction vertex is located inside the fiducial volume of the detector.  
For each event, the detector gives the topology as track-like (mainly charged-current interactions of muon \ns) or shower-like (all other types of interactions), the measured energy and the arrival direction. The latter has a  median angular error  which is of ${\mathcal O}(1^\circ)$ for track-like and ${\mathcal O}(10^\circ)$ for shower-like events \cite{Aartsen:2014gkd,Kopper:2015vzf,Kopper:2017zzm}. Out of a total of $N_\nu=79$ HESE events, 
58 are shower-like. It is also estimated that $N_b = 40.8\substack{+18.7 \\ -11.2}$  events are due to background (see \cite{Kopper:2017zzm} for details). 

The second set is the 6-year list of $N_\nu=29$ track-like events \cite{Aartsen:2016xlq}, with their vertex either inside or outside the fiducial volume. For this event sample, the detector field of view is restricted to the Northern Hemisphere.

In the main analysis of \aug\ (Table 1 in the \aug\ paper), 23 \sbg\ are considered. They were obtained by selecting, from the 3-year Fermi-LAT  catalog \cite{Ackermann:2012vca}, the objects that are closest (distance $d<250$ Mpc) and brightest (flux density larger than 0.3 Jy at frequency $f=1.4~{\rm GHz}$ \footnote{Throughout this paper, the symbol $f$ will be used for frequency, to be distinguished from $\nu$, the symbol of the neutrino particle. Also, note that $1~ {\rm Jy}= 10^{-26} {\rm W m^{-2} Hz^{-1}}$. }). 
 This list is  incomplete in the southern hemisphere (declination $\delta \lesssim -35^\circ$).
The authors of \aug\ examined other \sbg\ lists as well -- specifically those in the  Fermi-LAT Third Source Catalog \cite{Acero:2015hja} and in the catalog by Becker et al. \cite{Becker:2009hw}   (see Appendix \ref{app:catalogs}) --  to further corroborate the anisotropy result, which indeed turned out to be robust.  Still, however, all of the \sbg\ lists in \aug\ were incomplete to some extent, especially in the southern hemisphere\footnote{The situation is expected to vastly improve in the next decade with the new deep multi-color surveys by the Large Synoptic Survey Telescope (LSST) \cite{Ivezic:2008fe} in the South  and the nearby all-sky space missions Euclid \cite{Laureijs:2011gra}, WFIRST \cite{Dore:2018smn} and SPHEREx \cite{Dore:2018kgp}. }. 

Here we elaborate on the \aug\ approach of using \sbg\ from the Becker et al. catalog \cite{Becker:2009hw} by extending it to  the southern hemisphere. Specifically, the extension was obtained by a two-step process that involved extracting far infra-red observations of galaxies from the IRAS Revised Bright Galaxy sample \cite{Sanders:2003ms}, and combining them with radio data from the Australia-based HI Parkes All Sky Survey \cite{2001MNRAS.322..486B}.  Only sources with radio flux densities larger than $0.3$ Jy at $f=1.4$ GHz (to match the selection of \aug) were included in our final compilation (see Appendix \ref{app:catalogs} for more details).  The result of this selection process is a set of 45 \sbg, that are listed in Table \ref{tab:list}.  The set includes the 4 major contributors to the \aug\ anisotropy, and all but two \footnote{The two missing objects are MRK 231 and NGC3556.  The former is excluded by our selection criterion on the distance; the latter has flux density below the threshold we have imposed; see the Appendix.  }
of the \sbg\ that appear in Table 1 of \aug.  
 Fig. \ref{fig:results} (left panel) shows the distribution of the candidate sources and of the HESE \n\ data points in the sky, in Equatorial coordinates. The Galactic plane is shown as well. 

%%%%%%%%%%%%%%%%%%%%%%%%%%%%%%%%%%%%%
\subsection{Methodology}
\label{subsec:method}

 The statistical analysis is done using a likelihood ratio method, as outlined in \cite{Emig:2015dma}. Here the basics of the method will be summarized.  For each \n\ data point, $i$ ($i=1,2,...,N_\nu$), the statistical variable of interest is the normalized angular distance to the \emph{closest} candidate source: $r_i=Min_{j} (S_{ij}/\sigma_i)$. Here, $S_{ij}$ is the angular distance from the \n\ $i$ to the candidate $j$, and $\sigma_i$ is the angular error on the \n\ position (the error on the position of \sbg\ is negligible here).
   The aim of our analysis is to compare the distribution of $r_i$ for the data (the index $i$ will be dropped from now on, for brevity) with the predicted distributions for the null case, and for a case where a fraction $g$ of detected \ns\ truly is from the set of candidate sources considered (``true matches''). 
   The distribution of $r$ expected in the null case ($g=0$) can be found by calculating the probability $dP$ that, for each (fixed) neutrino in the data set, a value between $r$ and $r+dr$ is realized in the assumption that the positions of the candidate sources are uniformly distributed in the sky.  The calculation can be done analytically (the final result is somewhat involved, see \cite{Emig:2015dma}), or by Monte Carlo simulation. In the Monte Carlo simulation, the positions of the \n\ data points are kept fixed, and $K=10^6$ realizations of the positions of the candidate sources, drawn from the uniform distribution, are generated. For each realization, the histogram of the $r$-values is calculated, and the average over all realizations is finally obtained.  Note that this methodology is valid for sets of \n\ data with widely varying error sizes, like the HESE sample at hand, which is a mix of shower-like and track-like events.  We refer to \cite{Emig:2015dma} for more details on the method. 
   
   Regarding a case with a fraction of true matches, a useful reference is the ideal situation where all the \ns\ are from the candidate sources, $g=1$. In such case, to model the $r$-distribution it is necessary to know the statistical meaning of the angular errors on the data. This information is not known in detail from the \ic\ publications, therefore we are forced to make some assumptions. We will later verify that our main conclusions hold when some of these assumptions are relaxed.  In absence of deflection effects on the neutrino propagation (which is a valid assumption in the Standard Model of particle physics), and if the positional errors are Gaussian and sufficiently small, $r$ would follow a Rayleigh distribution \footnote{The Rayleigh distribution is not exactly valid here, because,  the positional errors are (for at least part of the data) not small, and not Gaussian \cite{Aartsen:2014gkd}.  
The use of bins of rather large width, $\Delta r=1$, should reduce the effect of deviations of the true distribution from the Rayleigh form. } (see, e.g., \cite{Windhorst84}): 
\begin{equation}
 R(r)=N_\nu \frac{r}{a^2} e^{-\frac{r^2}{2 a^2}} ~.
\label{rayleigh}
\end{equation}   
 Here $a$ is a constant that depends on the confidence level associated to the angular error. Motivated by the description of the errors in \cite{Aartsen:2014gkd,Kopper:2015vzf,Kopper:2017zzm}, we use the Rayleigh distribution with $a = (2 \ln 2)^{-1/2} \simeq 0.849$, corresponding to a 50\% confidence level.  
 
 In a realistic setup, with $g ={\mathcal O}(0.1)$, the total distribution of $r$ will be a linear combination of the null histogram and of $R(r)$, with weights $(1-g)$ and $g$ respectively. This intuitive result was checked using a Monte Carlo simulation.  In the simulation, $K=10^6$ realizations of the positions of the  candidate sources were generated from the uniform distribution.  For each realization, a set of $N_\mathrm{true}\simeq g N_\nu$ neutrino events were chosen randomly as true matches.  
For these, the $r$ values were drawn from the Rayleigh distribution.
 Otherwise, if a neutrino event was not chosen as a true match, its $r$ value was calculated in the same way as in the null case, as the angular distance to the nearest candidate source in the realization.
 \\
 
 When comparing the data with model predictions, two test statistics will be used: 
 
 \begin{itemize}
 
\item the number of positional coincidences, defined as the number $N_c$ of \n\ data points with $r<1$.   This is an intuitive, robust indicator and has been widely used in the literature over several decades (see, e.g., \cite{DeRuiter1977,Windhorst84}). 
 Its robustness is due to the fact that $N_c$ is expected to increase with  $g$ 
for all reasonable $r$-distributions of the true matches (although the rate of increase generally depends on the specific distribution).  This is not true, for example, for the number of data points in the bin $1<r<2$, whose behavior with varying $g$ is more model-dependent (it depends strongly on the $a$ parameter, for example).  Therefore, using $N_c$ as a test statistic is a conservative choice. 

\item the Kolmogorov-Smirnov (KS) test statistic (see, e.g., \cite{degroot2012probability}), defined as the supremum of the difference (in absolute value) of the cumulative probability distribution functions of the model and of the data: 
\begin{equation}
D_{KS} \equiv \sup_{j}  \left| {\rm CDF}(r_j)- {\rm CDF}(r) \right| ~. 
\label{dks}
\end{equation}
Note that $D_{KS}$ uses un-binned data, and is independent of the overall normalization of the model distribution.  
It is a global quantity, which can be used to estimate the overall compatibility of the $r$-distribution of the data with a given model.  Therefore, it is sensitive to several effects that could cause an incompatibility of data and model, other than the presence of true matches. In this respect, the two test statistics proposed here can be considered complementary. 

\end{itemize}

For each test statistic, $H$ ($H=N_c, D_{KS}$), the level of compatibility between data and model will be estimated using a $p$-value, ${\mathcal P}_H$. The latter is defined as the probability --- under the assumption that the model is true --- that a value of the indicator  is realized that is more discrepant from the mean, $\langle H  \rangle$, than the data, and in the same direction (suppression or excess) as the data. In other words\footnote{
This definition of $p$-value is a generalization of other definitions commonly encountered in the literature. For example, in \cite{Emig:2015dma} an excess in $N_c$ relative to the model was found, therefore only the case $H^{data} \geq \langle H \rangle$ was considered. 
}: 
\begin{equation}
{\mathcal P}_H \equiv \begin{cases} P(H \geq H^{data}) &\mbox{if } H^{data} \geq \langle H  \rangle \\ 
P(H \leq H^{data}) &\mbox{if } H^{data} <  \langle H  \rangle \end{cases} ~. 
\label{pvaluedef}
\end{equation}
In the following, the p-values for the two test statistics will be denoted as ${\mathcal P}_N$, and ${\mathcal P}_{KS}$. The case $H^{data}<\langle H  \rangle$ is only realized for $N_c$, where the value from the data is below the expectation of the model, as will be seen below. 
 A model will be considered disfavored if at least one  p-value falls below a minimum threshold $p_{min}=10^{-2}$.  

%%%%%%%%%%%%%%%%%%%%%%%%%%%%%%%%%%%%%
\section{Results}
\label{sec:res}

\begin{figure*}[t!]
\includegraphics[width=0.59\textwidth]{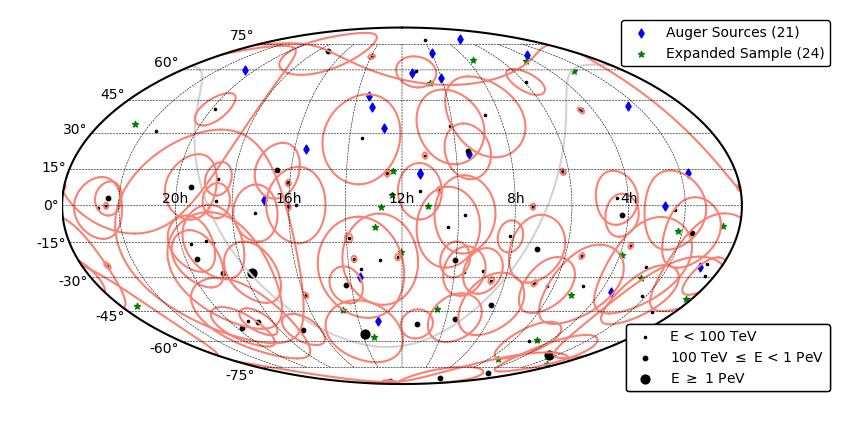} 
\hspace{0.4cm} 
\includegraphics[width=0.39\textwidth]{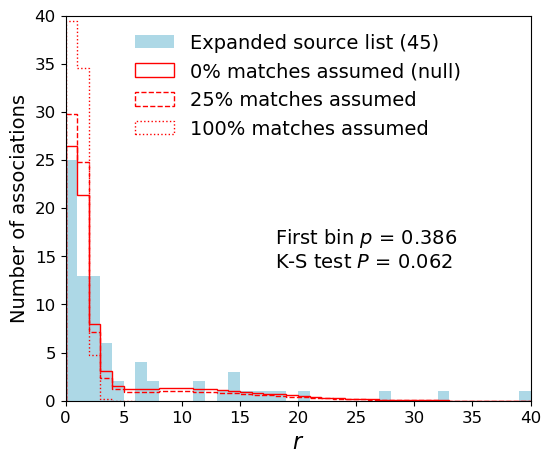}
 \caption{\label{fig:results} {\it Left:} Sky map (in J2000 equatorial coordinates) showing the arrival directions of the 6-year set of high-energy starting events (HESE) at the \ic\ \n\ observatory (black dots; with dot size indicating bins of observed energy, see legend) with their positional errors (ellipses). Also shown is the set of \sbg\ considered in the analysis. For these, different markers are used to distinguish the galaxies that appear in the primary analysis of the Pierre Auger collaboration paper (Table 1 in \cite{Aab:2018chp}).  {\it Right:} 
 The distribution of the minimum normalized angular distance, $r$, for the data (shaded) and for the null case (solid line). For illustration, the distributions for hypothetical cases with 25\% and 100\% assumed true matches (causally related to the candidate sources) are shown as well,
 see Sec. \ref{sec:data}. 
 }
 \end{figure*}
 
The main result of our analysis is shown in fig. \ref{fig:results} (right panel). 
It presents the distribution of $r$ for the HESE \n\ data and for the null case ($g=0$).  
For illustration, the extreme case with $g=1$, and an intermediate scenario with $g=0.25$ are shown as well. 

Let us first compare the data with the null hypothesis. 
 The number of observed coincidences is $N_c=25$, only slightly lower than the prediction in the null case.
  The corresponding $p$-value is ${\mathcal P}_N\simeq 0.39$.
Overall, the entire $r$-distribution is consistent with the one expected in the null case, with a slight tension in the second bin, where the data histogram is noticeably lower than the prediction.   We find  ${\mathcal P}_{KS}\simeq 0.062$, which is sufficiently large that the null hypothesis cannot be rejected.  These results lead  to the conclusion that there is \emph{no} indication of causal association of the HESE \ic\ \n\ data with the nearby \sbg\ in the considered sample.  We checked that a similar degree of consistency with the null case is found when restricting the analysis to the original \sbg\ list in \aug. Therefore, our conclusion is robust.   We  also find good agreement  with the null case when different bin sizes for the histogram in fig. \ref{fig:results} are used. For example, using a bin size $\Delta r=0.5 $, the number of data points in the first 4 bins are $N=12,13,6,7$, to be compared with $N\simeq 10.0, 16.5, 13.1, 8.3$ expected in the null hypothesis.  Again, the data is statistically consistent with the null hypothesis.

A relevant question is if, and to what degree, the data are consistent with a model with a small fraction of true matches. To answer this, we have repeated the analysis for models with $g>0$.    In agreement with intuition, and according to Eq. (\ref{rayleigh}), for increasing $g$ the predicted $r$-distribution becomes more strongly peaked at $r \sim 1$, with more power in first two bins, and the tail at $r>2$  becoming less populated.  Therefore, an increasing tension with the data is expected.  For example, for $g=0.1$ (i.e., 8 true matches)  the numbers of events expected in the first two bins are $N\simeq 27.8, 22.3$.  For $N_c$ the p-value is acceptable, ${\mathcal P}_{N}\simeq 0.26$.    However, the Kolmogorov-Smirnov p-value is ${\mathcal P}_{KS}\simeq 0.011$, which is only slightly above the threshold $p_{min}$. It reflects the disparity between data and model in the overall shape of the distribution, with the strongest tension being in the second bin.  
 For larger $g$, both p-values decrease; e.g., for $g=0.25$ (20 true matches, see Fig. \ref{fig:results}) we obtain ${\mathcal P}_{N}=0.13$, and ${\mathcal P}_{KS}\simeq 3.8 ~10^{-4}$.  Therefore this case is robustly disfavored by the KS test.  

 Our results for the Kolmogorov-Smirnov test  indicate that the largest amount of true matches that the data can tolerate is at the level of ten percent, $g \lesssim 0.1$. We stress that this conclusion is necessarily tentative, in the light of the cautionary statements made above about the meaning of the angular errors and the parameter $a$. Still, we find that even a drastically different value of $a$, e.g., $a=0.465$ (corresponding to 90\% confidence level errors), gives nearly identical values of ${\mathcal P}_{KS}$, indicating a robust result. If one relaxes the p-value threshold to $p_{min}=10^{-3}$, the KS test allows up to $16$ true matches ($g< 0.2$).  
Therefore, one reaches the broad conclusion that, conservatively, the largest acceptable value of $g$ lies between 0.1 and 0.2.

One may wonder if the sensitivity of the analysis suffers from considering a relatively large number of candidate sources, when in fact only 4 objects (in bold in Table \ref{tab:list}) were identified in \aug\ as accounting for most of the observed anisotropy. As a further test, we repeated the analysis using the latter subset. Results are, again, consistent with the null case: taking bins of width $\Delta r=1 $, the number of data points in the first 2 bins are
 $N=7,6$, whereas for the null hypothesis the prediction is $N\simeq 4.2, 9.2$. The two test statistics are consistent with the null case. 
   If 10\% of the events (8 true matches) were due to the 4 candidate sources,  then the predicted number of \n\ events in those bins would be  $N_c \simeq 7.7$ (for $0<r\leq1$)  and $N\simeq 11.7$ (for $1<r\leq 2$), which are in acceptable agreement with the observed counts, although with some tension in the second bin.  A p-value ${\mathcal P}_{KS}=0.13$ is found in this case.

In the same spirit of restricting the investigation to the (potentially) most relevant data, we have also repeated the analyses above for  subsets of the \n\ HESE data with higher observed energies: $E_{obs}>50,100,150$ TeV, corresponding to a number of \n\ events $N_\nu=57, 28,19$, respectively \footnote{Note how a large \n\ data set where many or most events are background can lead to a situation where the positional error ellipses cover most of the sky (see fig. \ref{fig:results}). In this case, one would find a large number of coincidences, that would make a true signal difficult to distinguish.  }.
Events with higher observed energy might be more likely to be of astrophysical origin, since atmospheric background fluxes are expected to be stronger at lower energy. In all cases, the results are consistent with the null hypothesis. 

Let us now discuss the analysis -- done with the same method outlined above -- for the set of track-like \n\ events from \cite{Aartsen:2016xlq} and the \sbg\ in Table \ref{tab:list}.  We find zero coincidences, $N_c=0$. The two minimum values of $r$ are $r=3.0, 5.7$, for the candidate sources NGC6240 (neutrino event number 6, observed energy $E_{obs}=770.0$ TeV) and Arp220 (neutrino event number 12, $E_{obs}=300.0$ TeV), respectively.  
These results are consistent with the null case, where the predicted numbers of events with $r<1$ and $0<r<6$ are $N\simeq 0.09, 2.0$ respectively. 
The KS test result is consistent with the null case as well, with p-value  ${\mathcal P}_{KS}\simeq 0.015$.
Because $r \geq 3$ for all the detected \ns, the possibility that any of them be causally related to the candidate sources is remote, 
and will not be considered further.  However, considering the low statistics of the track-like sample, it is intuitive that a (still undetected) contribution to the total \n\ flux from \sbg\ at the level of $\sim 10\%$ is allowed.  

%%%%%%%%%%%%%%%%%%%%%%%%%%%%%%%%%%%%%
\section{Discussion and conclusion}
\label{sec:disc}

To summarize, we find that there is \emph{no} indication of a causal correlation between the \n\ data in the two published \ic\ sets and the nearby starburst galaxies considered in \aug.  Due to the limited statistics, and to the large errors on the direction of the shower-like \n\ events,  a fraction of causally correlated 
HESE events in the detector at the level of $\sim 10\%$  -- which is approximately the size of the effect found in \aug\ -- is marginally allowed.

Our result is in agreement with previous multi-messenger studies, where the contribution of \sbg\ to the \ic\ \n\ data is constrained to be at the level of tens of percent or less (see, e.g., \cite{Bechtol:2015uqb}). More broadly, it is also consistent with two other arguments. 
The first is the existence of constraints on the \n\ luminosity of certain individual \sbg, obtained from their gamma-ray spectra \cite{Acero:2015hja,Acciari2009,Acero:2009nb}. Under naturalness assumptions, these spectra are expected to decline rapidly with energy and fall below the minimum required for detectability at IceCube.  
The second argument has to do with the ratio of local and cosmological contributions from the same class of sources.  If the \n\ production rate in \sbg\ tracks the star-formation rate, only $\sim 1\%$ or so of the total flux of \ns\ from \sbg\ can be from nearby galaxies ($z < 0.03$ in our sample, see the Appendix) , with the rest being diffuse, from sources at cosmological distances  (see, e.g., \cite{Ahlers:2014ioa,Emig:2015dma,Feyereisen:2016fzb}).  The local fraction can reach a maximum of $\sim 10\%$ for the most optimistic assumptions on the cosmological evolution of the \sbg\ population.  
We refer to \cite{Emig:2015dma} for a more extended discussion of these points. 

Future analyses with higher statistics \n\ data will be able to further constrain the allowed contribution of nearby \sbg. It is possible that the \aug\ anisotropy will be established  and confirmed to be due to \sbg, and at the same time the \n\ flux from the same sources will be constrained to a much smaller fraction. Such situation could be explained by the \n\ flux being mostly cosmological (whereas the short mean free path of the \uhs\ suppresses their cosmological flux), as we have discussed. It  may also favor scenarios with suppressed pion (and, therefore, \n) production efficiency, which can be realized depending on the properties of a galaxy (gas density, galactic wind, etc.), see, e.g., \cite{Chang:2014sua} for a discussion.  

In conclusion, the question of the  role of nearby starburst galaxies in the production of UHECR and \ns\ remains fairly open at this time. It is likely that significant advancements on this front will require disentangling contributions of several classes of objects to the \n\ flux, through extensive multi-year, multi-messenger campaigns that will lead us into the next decade. 

\begin{acknowledgments}
We thank the anonymous referee for insightful comments. 
CL is grateful to Lorenzo Perrone for useful discussions, and to Kohta Murase for early feedback. She acknowledges funding from the National Science Foundation grant number PHY-1613708. KLE acknowledges financial support from the Netherlands Organization for Scientific Research (NWO) through TOP grant 614.001.351. RAW is supported by NASA JWST grants NNX14AN10G and NAG5-12460.   
\end{acknowledgments}

\appendix
\section{Starburst galaxies catalogs and selection criteria}
\label{app:catalogs}

%\begin{array}{llll}
\begin{table}[tbp]
\centering
\begin{tabular}{|l|c|c|c|c|c|}
\hline
Name & RA (J2000) & Dec (J2000) & Distance (Mpc) & $S_{\mathrm{60 \,\mu m}}$ (Jy) & $S_{\mathrm{1.4\,GHz}}$ (Jy) \\
\hline
\hline
%%%%%%%%%%%%%%%%
GC0055$^{\ast}$ & 3.7664 & -39.2077 & 3.1 & 77.0 & 0.37  \\

NGC0157$^{\ast}$ & 8.6917 & -8.3981 & 21.92 & 17.93 & 0.31  \\

{\bf NGC0253} & 11.8776 & -25.2753 & 3.1 & 967.81 & 6.0  \\

SMC$^{\ast}$ & 13.2085 & -72.7876 & 0.06 & 6688.9 & 1.26  \\

NGC0660 & 25.7179 & 13.6358 & 12.33 & 65.52 & 0.37  \\

NGC0839$^{\ast}$ & 32.4288 & -10.1842 & 51.1 & 11.67 & 0.37  \\

NGC891 & 35.6392 & 42.3491 & 8.57 & 66.46 & 0.7  \\

Maffei2 & 40.4795 & 59.6041 & 3.32 & 135.0 & 1.01  \\

{\bf NGC1068} & 40.6645 & -0.0020 & 13.7 & 196.37 & 4.85  \\

NGC1097 & 41.6007 & -30.2717 & 16.8 & 53.35 & 0.41   \\

NGC1365 & 53.3839 & -36.1408 & 17.93 & 94.31 & 0.53   \\

IC342 & 56.7021 & 68.0961 & 4.6 & 180.8 & 2.25   \\

NGC1482$^{\ast}$ & 58.6658 & -20.5019 & 25.09 & 33.36 & 0.31   \\

NGC1569 & 67.7044 & 64.8479 & 4.6 & 54.36 & 0.4   \\

NGC1672 & 71.4279 & -59.2467 & 16.82 & 41.21 & 0.45   \\

NGC1808 & 76.9319 & -37.5228 & 12.61 & 105.55 & 0.5   \\

LMC$^{\ast}$ & 80.8938 & -69.7561 & 0.05 & 82917.0 & 1.21   \\

NGC2146 & 94.6571 & 78.3570 & 16.47 & 146.69 & 1.09   \\

NGC2403 & 114.2140 & 65.6026 & 3.22 & 41.47 & 0.39   \\

NGC2903 & 143.0460 & 21.5101 & 8.26 & 60.54 & 0.44   \\

NGC3034(M82) & 148.9680 & 69.6797 & 3.63 & 1480.42 & 7.29   \\

NGC3079 & 150.4910 & 55.6797 & 18.19 & 50.67 & 0.82   \\

NGC3256 & 156.9876 & -43.9090 & 35.35 & 102.63 & 0.64  \\

NGC3310 & 159.6910 & 53.5034 & 19.81 & 34.56 & 0.42  \\

NGC3521$^{\ast}$ & 166.4550 & 0.0375 & 6.84 & 49.19 & 0.35  \\

NGC3628 & 170.0818 & 13.6037 & 10.04 & 54.8 & 0.47  \\

NGC3627 & 170.0857 & 13.0005 & 10.04 & 66.31 & 0.46  \\

NGC3690 & 172.1340 & 58.5622 & 47.74 & 113.05 & 0.66  \\

NGC4038/9$^{\ast}$ & 180.4873 & -18.8984 & 21.54 & 45.16 & 0.54  \\

NGC4254$^{\ast}$ & 184.7063 & 14.4272 & 15.29 & 37.46 & 0.37  \\

NGC4303 & 185.4808 & 4.4733 & 15.29 & 37.27 & 0.44  \\

NGC4631 & 190.5330 & 32.5420 & 7.73 & 85.4 & 1.12  \\

NGC4666 & 191.2860 & -0.4619 & 12.82 & 37.11 & 0.43  \\

NGC4818$^{\ast}$ & 194.2083 & -8.5272 & 9.37 & 20.12 & 0.45  \\

{\bf NGC4945} & 196.3792 & -49.4544 & 3.92 & 625.46 & 6.6  \\

NGC5055(M63) & 198.9560 & 42.0293 & 7.96 & 40.0 & 0.35  \\

ESO173-G015$^{\ast}$ & 201.8517 & -57.4900 & 32.44 & 81.44 & 0.48  \\

NGC5194(M51) & 202.4700 & 47.1952 & 8.73 & 97.42 & 1.31  \\

{\bf NGC5236(M83)} & 204.2532 & -29.8586 & 3.6 & 265.84 & 2.44  \\

NGC5643$^{\ast}$ & 218.2197 & -44.1990 & 13.86 & 23.48 & 0.36  \\

UGC09913(Arp220) & 233.7379 & 23.5028 & 79.9 & 104.09 & 0.32  \\

NGC6240 & 253.2442 & 2.4008 & 103.86 & 22.94 & 0.65  \\

NGC6946 & 308.7180 & 60.1539 & 5.32 & 129.78 & 1.4  \\

NGC7331 & 339.2670 & 34.4156 & 14.71 & 45.0 & 0.37  \\

NGC7582$^{\ast}$ & 349.5925 & -42.3719 & 21.29 & 52.2 & 0.68  \\

\hline
\end{tabular}
\caption{The starburst galaxies used in this work with their Equatorial coordinates (in degrees) and their flux densities at 1.4 GHz and $60 \mu m$.  The flux extracted from CHIPASS has a maximum error of $\sim 20\%$. Asterisks mark the galaxies of our Set 2, which were not included in the \aug\ analysis.  The four \sbg\ that contribute the most to the \aug\ anisotropy are marked in bold.
}
\label{tab:list}
\end{table}

In this appendix, details are given on how the list of \sbg\ considered here (Table \ref{tab:list}) was obtained. 
For comparison, let us first summarize the approach and data in the \aug\ paper. The primary result there was obtained using the Fermi SBG catalog, but other data sets were also tested with results (i.e.,  a significant anisotropy) consistent with the primary analysis. Below the different \sbg\ sets are described briefly. They are:

\begin{itemize}

\item  A selection of \sbg\ searched for in gamma-ray emission with Fermi-LAT \cite{Ackermann:2012vca}. In the search, a set of SBGs was compiled from a survey of the dense molecular gas tracer, HCN \cite{Gao:2003qp}.   This HCN survey is statistically complete for northern galaxies (declination $\delta \geq -35^\circ$) with flux density at far-infrared (FIR) wavelength $\lambda=100{\rm \mu m}$ of $S_{100\mu m} \geq 100$ Jy (corresponding to, approximately, a flux density  $S_{60{\rm \mu m}} > 50$ Jy at $\lambda=60 {\rm \mu m}$).  However, the full \sbg\ set includes additional, fainter \sbg\ ($S_{60{\rm \mu m}} < 50$ Jy), which were used to establish the relationship between HCN luminosities and star formation rates over a wide range of FIR luminosities (see \cite{Gao:2003qp} for more details).  Therefore, the full data set is not complete, and it does not  fully satisfy the assumption of a uniformly sampled all-sky distribution (which is required by our method of analysis).

\item The Fermi-LAT Third Source Catalog (3FGL) \cite{Acero:2015hja}, from which the list of the 6 \sbg\ observed in gamma rays (NGC 253, M82, NGC 4945, NGC 1068, Circinus\footnote{Circinus  is not included in the list of \sbg\  that result from our selection criteria (discussed below, see Table \ref{tab:list}). This might due to Circinus being located within $10^\circ$ from the Galactic Plane, where FIR and radio data are harder to obtain reliably due to Galactic foreground confusion. }, NGC 2146) was obtained.

\item The 2009 catalog by Becker et al. \cite{Becker:2009hw}.  These authors use the FIR flux as a proxy for star-formation. Certain criteria were required for inclusion in the catalog, to ensure its completeness and to remove contamination from other astrophysical objects. They are: \\
(1) a FIR flux density of $S_{60{\rm \mu m}} > 4$ Jy, \\
 (2) a radio flux density of $S_{\rm 1.4GHz} > 20$ mJy, \\
  (3) a ratio of FIR to radio flux densities of $S_{60{\rm \mu m}} / S_{\rm 1.4GHz}  > 30$.  \\
  (4) An additional constraint on the redshift of $z <0.03$ (distance $D<130$ Mpc), was placed to ensure that the \sbg\ were locally within the Super-Galactic plane \cite{1953AJ.....58...30D,Lahav:1998mm}.  \\
The radio fluxes were extracted from the NRAO VLA Sky Survey (NVSS) \cite{Condon:1998iy}, which is limited in declination to $\delta > -40^\circ$.  Therefore, the Becker et al. compilation has a similar restriction in declination. 
 It differs from Fermi \sbg\ set largely in its distance requirement and its all-sky completeness.

\end{itemize}

We took a fresh look at the problem of producing a sample of candidate \sbg\ as detailed and complete as possible.
While the majority of IceCube neutrino events fall within the southern hemisphere, current \sbg\ catalogs are missing coverage in the crucial declinations of $\delta < - 40^\circ$,  due to a lack of radio surveys in that region (e.g. in HCN, or in the 1.4 GHz continuum).  With the motivation to correct for this shortage, we turned to the Continuum HI Parkes All Sky Survey (CHIPASS) to extract the flux densities at  $f=1.4 ~{\rm GHz}$ for the brightest \sbg\ located at the most southern declinations. 
Essentially, 
we followed the same procedure that led to the Becker et al. catalog, 
with the additional 1.4 GHz measurements for sources with $S_{\rm 1.4GHz} > 0.3$ Jy from the CHIPASS. The goal here is to extend the galaxy sample that was used in AUG to include sources of $\delta < - 40^\circ$.

Here we briefly introduce the CHIPASS. It is a survey covering the Equatorial and Southern sky at $\delta < +25^\circ$  with the Parkes telescope, a single dish of 64 m in diameter located in Australia. The relatively poor angular resolution (14.4$'$ at 1.4 GHz) results in the image sensitivity being limited by confusion noise of $\sigma_c = 0.03$ Jy/beam. Hence, we expect our flux measurements to be somewhat beam-diluted, since the angular sizes of nearby galaxies are roughly 10$'$ or less. However, the brightest galaxies, presumably contributing most to the neutrino and/or UHECR flux, would not be much affected by this dilution. Furthermore, by employing criterium (3) above, we strongly reduce the possibility of e.g. chance radio galaxies contaminating the radio flux of the sources. 

As a first step towards creating a new set of \sbg, we used the IRAS Revised Bright Galaxy sample to identify the galaxies that are brightest at FIR frequencies.  We imposed the same condition as in \cite{Becker:2009hw} on the FIR flux density, see criterion (1) above.   In this way, a set of 195 \sbg\ (Set 1) was obtained.  
For each candidate  galaxy in Set 1, we reprojected the CHIPASS image  \cite{Calabretta:2013yka} on a 15$^\circ \times 15^\circ$ region centered at the object's coordinates.  We then integrated the 1.4 GHz flux density within a circular aperture of width two times the beam full-width at half-maximum (i.e. 28.8$'$). Instead of the criterion (2) above on the flux density, a stronger condition was required, namely  $S_{1.4GHz}  >0.3$ Jy at $f=1.4$ GHz to match the one used in \aug.  Finally, the conditions (3) and (4) as in \cite{Becker:2009hw} were 
imposed. 
 The resulting list, Set 2, contained 13 \sbg.  Finally, the union of Set 2 and the corresponding selection from the Becker et al. catalog was taken, producing the final list of 45 starburst galaxies shown in Table \ref{tab:list}.

% Create the reference section using BibTeX:
%%%%%%%%%%%%%%%%%%%%%%%%%%%%%%

\bibliographystyle{JHEP}
\bibliography{refs}

\end{document}